\input harvmac.tex
\input graphicx

\yearltd=\year 

\def\cN{{\cal N}}

\def\cite{{\bf CITE}}  
  
\lref\HoweQZ{
  P.~S.~Howe and K.~S.~Stelle,
  ``The Ultraviolet Properties Of Supersymmetric Field Theories,''
  Int.\ J.\ Mod.\ Phys.\  A {\bf 4}, 1871 (1989).
}

\lref\BohrVanhove{N.~E.~J.~Bjerrum-Bohr and Pierre~Vanhove, work in progress
}

\lref\CachazoGA{
  F.~Cachazo and P.~Svrcek,
  ``Lectures on twistor strings and perturbative Yang-Mills theory,''
  PoS {\bf RTN2005}, 004 (2005)
  [arXiv:hep-th/0504194].
}

\lref\BernAN{
  Z.~Bern and D.~C.~Dunbar,
  ``A Mapping between Feynman and string motivated one loop rules in gauge
  theories,''
  Nucl.\ Phys.\  B {\bf 379}, 562 (1992).
}

\lref\gvstringloop{
M.B. Green and  P. Vanhove,
``The low energy
expansion of the one-loop type II superstring amplitude,'' Phys. Rev.
{\bf D 61}, 104011 (2000) [arXiv:hep-th/9910056].
}

\lref\ArkaniHamedYF{
  N.~Arkani-Hamed and J.~Kaplan,
  ``On Tree Amplitudes in Gauge Theory and Gravity,''
  arXiv:0801.2385 [hep-th].
}

\lref\Treegravity{
F.~Cachazo and P.~Svr\v{c}ek,
``Tree level recursion relations in general relativity,''
hep-th/0502160.\hfill\break
J.~Bedford, A.~Brandhuber, B.~Spence and G.~Travaglini,
``A recursion relation for gravity amplitudes,''
Nucl.\ Phys.\ B {\bf 721}, 98 (2005) [hep-th/0502146].\hfill\break
P.~Benincasa, C.~Boucher-Veronneau and F.~Cachazo, ``Taming tree
amplitudes in general relativity,'' JHEP {\bf 0711}, 057 (2007)
[arXiv:hep-th/0702032].\hfill\break
H.~Elvang and D.~Z.~Freedman, ``Note on graviton MHV amplitudes,''
arXiv:0710.1270 [hep-th].
}

\lref\PeetersUB{
K.~Peeters, P.~Vanhove and A.~Westerberg,
``Chiral splitting and world-sheet gravitinos in higher-derivative string amplitudes,''
Class.\ Quant.\ Grav.\  {\bf 19}, 2699 (2002)
[arXiv:hep-th/0112157].
}

\lref\GreenZZB{
M.~B.~Green, H.~Ooguri and J.~H.~Schwarz,
``Decoupling Supergravity from the Superstring,''
Phys.\ Rev.\ Lett.\  {\bf 99}, 041601 (2007)
[arXiv:0704.0777 [hep-th]].
}

\lref\MinahanHA{
J.~A.~Minahan,
``One Loop Amplitudes On Orbifolds And The Renormalization Of Coupling
Constants,''
Nucl.\ Phys.\  B {\bf 298}, 36 (1988).
}

\lref\BernPK{
Z.~Bern and D.~A.~Kosower,
``Absence Of Wave Function Renormalization In Polyakov Amplitudes,''
Nucl.\ Phys.\  B {\bf 321}, 605 (1989).
}

\lref\NastiSR{A.~Nasti and G.~Travaglini,
``One-loop N=8 Supergravity Amplitudes from MHV Diagrams,''
arXiv:0706.0976 [hep-th].
}

\lref\BerkovitsVC{ N.~Berkovits, ``New higher-derivative R**4
theorems,'' Phys.\ Rev.\ Lett.\  {\bf 98}, 211601 (2007)
[arXiv:hep-th/0609006].\hfill\break
N.~Berkovits and N.~Nekrasov, ``Multiloop superstring amplitudes
from non-minimal pure spinor formalism,'' JHEP {\bf 0612}, 029
(2006) [arXiv:hep-th/0609012].
}

\lref\GreenGT{
M.~B.~Green, J.~G.~Russo and P.~Vanhove,
``Non-renormalisation conditions in type II string theory and maximal
supergravity,''
JHEP {\bf 0702}, 099 (2007)
[arXiv:hep-th/0610299].
}

\lref\BjerrumBohrYW{
N.~E.~J.~Bjerrum-Bohr, D.~C.~Dunbar, H.~Ita, W.~B.~Perkins and K.~Risager,
``The no-triangle hypothesis for N = 8 supergravity,''
JHEP {\bf 0612}, 072 (2006)
[arXiv:hep-th/0610043].
}

\lref\BernBB{
Z.~Bern, N.~E.~J.~Bjerrum-Bohr and D.~C.~Dunbar,
``Inherited twistor-space structure of gravity loop amplitudes,''
JHEP {\bf 0505}, 056 (2005)
[arXiv:hep-th/0501137].
}

\lref\BjerrumBohrJR{
N.~E.~J.~Bjerrum-Bohr, D.~C.~Dunbar, H.~Ita, W.~B.~Perkins and K.~Risager,
JHEP {\bf 0601}, 009 (2006)
[arXiv:hep-th/0509016].
}

\lref\BernKR{
Z.~Bern, L.~J.~Dixon and D.~A.~Kosower,
``Dimensionally regulated pentagon integrals,''
Nucl.\ Phys.\  B {\bf 412}, 751 (1994)
[arXiv:hep-ph/9306240].
}

\lref\BernEM{
Z.~Bern, L.~J.~Dixon and D.~A.~Kosower,
``Dimensionally Regulated One Loop Integrals,''
Phys.\ Lett.\  B {\bf 302}, 299 (1993)
[Erratum-ibid.\  B {\bf 318}, 649 (1993)]
[arXiv:hep-ph/9212308].
}

\lref\BernML{Z.~Bern, L.~J.~Dixon, M.~Perelstein and J.~S.~Rozowsky, ``Multi-leg
one-loop gravity amplitudes from gauge theory,'' Nucl.\ Phys.\  B
{\bf 546}, 423 (1999) [arXiv:hep-th/9811140].
}

\lref\BernSV{ Z.~Bern, L.~J.~Dixon, D.~C.~Dunbar, M.~Perelstein and
J.~S.~Rozowsky, ``On the relationship between Yang-Mills theory and
gravity and its implication for ultraviolet divergences,'' Nucl.\
Phys.\ B {\bf 530}, 401 (1998) [hep-th/9802162],\hfill\break
%
Z.~Bern, L.~J.~Dixon, M.~Perelstein, D.~C.~Dunbar and
J.~S.~Rozowsky, ``Perturbative relations between gravity and gauge
theory,'' Class.\ Quant.\ Grav.\  {\bf 17}, 979 (2000)
[hep-th/9911194].
}

\lref\SpinorHelicity{Z.~Xu, D.~H.~Zhang and L.~Chang,
``Helicity Amplitudes For Multiple Bremsstrahlung In Massless Nonabelian Gauge
Theories,''
Nucl.\ Phys.\ B {\bf 291}, 392 (1987).
}

\lref\BernXJ{
Z.~Bern, J.~J.~Carrasco, D.~Forde, H.~Ita and H.~Johansson,
``Unexpected Cancellations in Gravity Theories,''
arXiv:0707.1035 [hep-th].
}

\lref\GreenUJ{
M.~B.~Green, J.~G.~Russo and P.~Vanhove,
``Low energy expansion of the four-particle genus-one amplitude in type II
superstring theory,''
arXiv:0801.0322 [hep-th].
}

\lref\BjerrumBohrXX{
N.~E.~J.~Bjerrum-Bohr, D.~C.~Dunbar and H.~Ita,
``Six-point one-loop N = 8 supergravity NMHV amplitudes and their IR behaviour,''
Phys.\ Lett.\  B {\bf 621}, 183 (2005)
[arXiv:hep-th/0503102].
}

\lref\NequalEight{ E.~Cremmer, B.~Julia and J.~Scherk,
``Supergravity theory in 11 dimensions,''
Phys.\ Lett.\ B {\bf 76} (1978) 409;\hfill\break
E.~Cremmer and B.~Julia,
``The N=8 Supergravity Theory. 1. The Lagrangian,''
Phys.\ Lett.\ B {\bf 80} (1978) 48.}

\lref\PolchinskiRR{
J.~Polchinski,
``String theory. Vol. 2: Superstring theory and beyond,''
{\it  Cambridge, UK: Univ. Pr. (1998) 531 p}
}

\lref\GreenSP{
M.~B.~Green, J.~H.~Schwarz and E.~Witten,
``Superstring Theory. Vol. 2: Introduction,''
{\it  Cambridge, Uk: Univ. Pr. ( 1987) 469 P. ( Cambridge Monographs
On Mathematical Physics)}}

\lref\DHokerTA{
E.~D'Hoker and D.~H.~Phong,
``The Geometry of String Perturbation Theory,''
Rev.\ Mod.\ Phys.\  {\bf 60}, 917 (1988).
}

\lref\HoweStelle{ P.~S.~Howe and K.~S.~Stelle,
``Supersymmetry counterterms revisited,''
Phys.\ Lett.\ B {\bf 554}, 190 (2003) [hep-th/0211279].
}

\lref\HoweTH{
P.~S.~Howe and U.~Lindstrom,
``Higher Order Invariants In Extended Supergravity,''
Nucl.\ Phys.\  B {\bf 181}, 487 (1981).
}

\lref\KalloshFI{
R.~E.~Kallosh,
``Counterterms in extended supergravities,''
Phys.\ Lett.\  B {\bf 99}, 122 (1981).
}

\lref\KalloshYM{
R.~Kallosh,
``The Effective Action of N=8 Supergravity,''
arXiv:0711.2108 [hep-th].
}

\lref\GreenYU{
M.~B.~Green, J.~G.~Russo and P.~Vanhove,
``Ultraviolet properties of maximal supergravity,''
Phys.\ Rev.\ Lett.\  {\bf 98}, 131602 (2007)
[arXiv:hep-th/0611273].
}

\lref\Kawaixq{
H.~Kawai, D.~C.~Lewellen and S.~H.~H.~Tye,
``A Relation Between Tree Amplitudes Of Closed And Open Strings,''
Nucl.\ Phys.\  B {\bf 269}, 1 (1986).}

\lref\SannanTZ{S.~Sannan,
``Gravity As The Limit Of The Type II Superstring Theory,''
Phys.\ Rev.\  D {\bf 34}, 1749 (1986).
}

\lref\BernJI{
Z.~Bern and A.~K.~Grant,
``Perturbative gravity from {QCD} amplitudes,''
Phys.\ Lett.\  B {\bf 457}, 23 (1999)
[arXiv:hep-th/9904026].\hfill\break
S.~Ananth and S.~Theisen,
``KLT relations from the Einstein-Hilbert Lagrangian,''
Phys.\ Lett.\  B {\bf 652}, 128 (2007)
[arXiv:0706.1778 [hep-th]].
}

\lref\BernKJ{
Z.~Bern,
``Perturbative quantum gravity and its relation to gauge theory,''
Living Rev.\ Rel.\  {\bf 5}, 5 (2002)
[arXiv:gr-qc/0206071].
}

\lref\EffKLT{Z.~Bern, A.~De Freitas and H.~L.~Wong,
``On the coupling of gravitons to matter,''
Phys.\ Rev.\ Lett.\  {\bf 84}, 3531 (2000) [hep-th/9912033].\hfill\break
N.~E.~J.~Bjerrum-Bohr,
``String theory and the mapping of gravity into gauge theory,''
Phys.\ Lett.\ B {\bf 560}, 98 (2003) [hep-th/0302131];
``Generalized string theory mapping relations between gravity and gauge
theory,''
Nucl.\ Phys.\ B {\bf 673}, 41 (2003) [hep-th/0305062];\hfill\break
N.~E.~J.~Bjerrum-Bohr and K.~Risager,
``String theory and the KLT-relations between gravity and gauge theory
including external matter,''
Phys.\ Rev.\ D {\bf 70}, 086011 (2004) [hep-th/0407085].
}

\lref\BerGiKu{ F.~A.~Berends, W.~T.~Giele and H.~Kuijf,
``On Relations Between Multi - Gluon And Multigraviton Scattering,''
Phys.\ Lett.\ B {\bf 211}, 91 (1988).
}

\lref\BernDW{
Z.~Bern, L.~J.~Dixon and D.~A.~Kosower,
``On-Shell Methods in Perturbative QCD,''
Annals Phys.\  {\bf 322}, 1587 (2007)
[arXiv:0704.2798 [hep-ph]].
}

\lref\WittenTopologicalString{
E.~Witten,
``Perturbative gauge theory as a string theory in twistor space,''
Commun.\ Math.\ Phys.\  {\bf 252}, 189 (2004)
[hep-th/0312171].}

\lref\Bernhh{
Z.~Bern, J.~J.~Carrasco, L.~J.~Dixon, H.~Johansson, D.~A.~Kosower and R.~Roiban,
``Three-Loop Superfiniteness of N=8 Supergravity,''
Phys.\ Rev.\ Lett.\  {\bf 98}, 161303 (2007)
[arXiv:hep-th/0702112].
}

\lref\BernKD{
Z.~Bern, L.~J.~Dixon and R.~Roiban,
``Is N = 8 supergravity ultraviolet finite?,''
Phys.\ Lett.\  B {\bf 644}, 265 (2007)
[arXiv:hep-th/0611086].
}

\lref\GreenSW{
M.~B.~Green, J.~H.~Schwarz and L.~Brink,
``N=4 Yang-Mills And N=8 Supergravity As Limits Of String Theories,''
Nucl.\ Phys.\  B {\bf 198}, 474 (1982).
}

\lref\StringBased{
Z.~Bern and D.~A.~Kosower,
``Efficient Calculation Of One Loop QCD Amplitudes,''
Phys.\ Rev.\ Lett.\  {\bf 66}, 1669 (1991);
``The Computation of loop amplitudes in gauge theories,''
Nucl.\ Phys.\ B {\bf 379}, 451 (1992);\hfill\break
Z.~Bern,
``A Compact representation of the one loop N gluon amplitude,''
Phys.\ Lett.\ B {\bf 296}, 85 (1992).\hfill\break
Z.~Bern, D.~C.~Dunbar and T.~Shimada,
``String based methods in perturbative gravity,''
Phys.\ Lett.\ B {\bf 312}, 277 (1993) [hep-th/9307001].\hfill\break
D.~C.~Dunbar and P.~S.~Norridge,
``Calculation of graviton scattering amplitudes using string based methods,''
Nucl.\ Phys.\ B {\bf 433}, 181 (1995) [hep-th/9408014];\hfill\break
%
D.~C.~Dunbar and N.~W.~P.~Turner,
``Gravity and form scattering and renormalisation of gravity in six and eight
Class.\ Quant.\ Grav.\  {\bf 20}, 2293 (2003) [hep-th/0212160];\hfill\break
%
D.~C.~Dunbar and P.~S.~Norridge,
``Infinities within graviton scattering amplitudes,''
Class.\ Quant.\ Grav.\  {\bf 14} (1997) 351 [hep-th/9512084].
}

\hoffset 0truecm \voffset -0.2truecm


\def\sg(#1){{\rm sign}(#1)}

\def\ap{{\alpha'}}
\def\<#1>{\langle #1\rangle}

\def\eqd#1#2{\xdef #1{(\secsym\the\meqno)}\writedef{#1\leftbracket#1}%
\global\advance\meqno by1$$\displaystyle #2\eqno#1\eqlabeL#1$$}
\noblackbox
\baselineskip 14pt plus 2pt minus 2pt
\Title{\vbox{\baselineskip12pt
\hbox{IPhT-T-08/018}}}{\vbox{
\centerline{Explicit Cancellation of Triangles}
\centerline{in One-loop Gravity  Amplitudes}}}
\centerline{N. E. J. Bjerrum-Bohr${}^1$ and Pierre Vanhove${}^{2,3}$ }
\medskip\medskip
\centerline{${}^1$ \sl School of Natural Sciences,}
\centerline{\sl Institute for Advanced Study,}
\centerline{\sl Einstein Drive, Princeton,}
\centerline{\sl NJ 08540, USA}
\medskip\smallskip
\centerline{${}^2$ \sl Institut de Physique Th\'eorique CEA,}
\centerline{\sl Orme des merisiers,}
\centerline{\sl F-91191 Gif-sur-Yvette, France}
\medskip\smallskip
\centerline{${}^3$ \sl Niels Bohr Institute,}
\centerline{\sl University of Copenhagen,}
\centerline{\sl Blegdamsvej 17, Copenhagen \O,}
\centerline{\sl DK--2100, Denmark}
\bigskip\bigskip \centerline{\tt
email: bjbohr@ias.edu,\    pierre.vanhove@cea.fr}
\bigskip\bigskip\bigskip
\centerline{\bf  Abstract} We analyse one-loop graviton amplitudes
in the field theory limit of a genus-one string theory computation.
The considered amplitudes can be dimensionally reduced to lower
dimensions preserving maximal supersymmetry. The particular case of
the one-loop five-graviton amplitude is worked out in detail and
explicitly features no triangle contributions.
Based on a recursive form of the one-loop amplitude we investigate
the contributions that will occur at $n$-point order in relation to
the ``no-triangle'' hypothesis  of ${\cal N} = 8$ supergravity.
We argue that the origin of unexpected cancellations observed in
gravity scattering amplitudes is linked to general coordinate
invariance of the gravitational action and the summation over all
orderings of external legs. Such cancellations are instrumental in
the extraordinary good ultra-violet behaviour of $\cN=8$
supergravity amplitudes and will play a central role in improving
the high-energy behaviour of gravity amplitudes at more than one
loop.

\Date{}
\nfig\figred{}
\nfig\figfive{}
\nfig\figcontractions{}
\newsec{Introduction}

Explicit evaluation of graviton scattering amplitudes is a complex
and difficult subject using traditional Feynman diagram techniques.
Amplitudes for trees and loops with unspecified external polarisation
tensors tend to be rather unmanageable, to hide manifest symmetries
and to exhibit undesirable features such as a factorial increase
in complexity with the number of external legs.
This makes the current knowledge of perturbative scattering amplitudes
for gravity limited and to a large degree based on assumptions from
power counting arguments rather than explicit calculations.
In the context of four dimensional maximal supergravity
power counting arguments indicate possible ultra-violet
divergences at three loops~\refs{\HoweQZ,\HoweStelle}, at
seven loops~\refs{\HoweTH}, at eight
loops~\refs{\KalloshFI,\KalloshYM}
or at nine loops~\refs{\GreenYU} depending on the implemented
superspace formalism. But so far no divergences have been
found in explicit calculations~\refs{\Bernhh}.

To avoid the myriad of tensor contractions and to generally simplify
calculations associated  with a  conventional field theory approach,
string theory can be used as  a guideline for  calculations.
Expressions for field theory amplitudes preserving supersymmetry
can  be derived  in the  infinite tension  limit
($\alpha'\rightarrow  0$) of  the string. String  theory  rules for
graviton amplitudes that  hold at tree level have  been formulated
very elegantly   by
Kawai, Lewellen and Tye~\refs{\Kawaixq} and in~\refs{\BerGiKu}.
Graviton amplitudes at tree level
from string theory was also investigated in~\refs{\SannanTZ}.
Interestingly  such rules also hold in a number
of different scenarios~\refs{\BernJI,\BernKJ} with
various matter contents~\refs{\EffKLT}. At
one-loop level  string   based  rules   have  been  formulated   for
amplitude calculations in both gauge theory
and gravity~\refs{\GreenSW,\StringBased}. This paper
investigates perturbative scattering amplitudes for gravitons
in maximal supergravity at one-loop using string theory based
techniques relying on the RNS formalism~\refs{\DHokerTA,\GreenSP,
\PolchinskiRR}.  In  this work we will consider the conventional
field theory limit of a one-loop string amplitude and  we will
not be affected by the issue raised in~\refs{\GreenZZB}.

In $D$ dimensions due to the two derivative coupling nature
of gravitational interactions an $n$-graviton amplitude at
one loop in gravity has the mass dimension
\eqn\eAnUV{ [{\cal M}_{n}]\sim  mass^{D}}
The dimensionful coupling of graviton amplitudes will render
gravity inherently non-renormalisable and the $n$-point one-loop
pure graviton amplitude is na\"{\i}vely given by a Feynman
integral with $2n$ powers of loop momenta in the numerator
\eqd\eIgrav{{\cal M}_{n}\sim\int d^{D}\ell \,{
\prod_{j=1}^{2n} \ell\cdot q_{j}\over \prod_{i=1}^n
(\ell-k_{1\cdots i})^2} }
Here $k_{1\cdots i}=k_{1}+\cdots +k_{i}$, and in the numerator
$q_{j}$ represents some (linear combination) of the external
momenta. Using this count for amplitudes in maximally
supergravity~\refs{\NequalEight} the maximum number of loop momenta
expected to be in the numerator in this case is reduced by eight by
supersymmetry. This leads to an overall total number of $2n-8$
powers of loop momenta in the numerator. This mean for ${\cal N} =
8$ graviton amplitudes in $D = 4$ that we are expected to observe
triangle integral functions at five points and both triangle and
bubble integral functions for amplitudes with six external legs.
For seven and higher point amplitudes besides triangle and bubble
integral functions - non-analytic rational contributions should be
present as well in the amplitude.

Recently, initiated by a paper by Witten~\refs{\WittenTopologicalString},
there has been  new explicit
calculations of scattering amplitudes, both for gauge theories (for
a review see~\refs{\BernDW,\CachazoGA}) and gravity.
These new results are to a large extend based on the
spinor-helicity~\refs{\SpinorHelicity} formalism in $D=4$.
This has led to new information about scattering
amplitudes for gravity and has allowed power counting estimates for
graviton amplitudes to be tested by explicit computation.
In the theory of maximal supergravity it
has been observed in a number of concrete amplitude
computations~\refs{\BernSV,\BernBB,\BjerrumBohrXX,\BjerrumBohrYW,
\BernXJ} that one-loop amplitudes exhibits mysterious unexpected
simplifications. These simplifications renders the integral
functions in gravity closer to Yang-Mills theory than would
otherwise be expected from the na\"{\i}ve counting that was
presented above. This has also been referred to as the
``no-triangle'' hypothesis of ${\cal N} = 8$
supergravity~\refs{\BjerrumBohrYW,\BernXJ}.

The concept of  unexpected simplifications is also
supported in a number of recent string theory
computations~\refs{\GreenYU,\GreenGT} where important input from the
pure spinor formalism of Berkovits~\refs{\BerkovitsVC} and string
theory dualities points towards a much better UV-behaviour for gravity
amplitudes than one should expect from power counting in
supersymmetry alone.

In concrete computations of ${\cal N}=8$ supergravity amplitudes
at most $n-4$ powers of loop momenta appear to be present in the
numerator of a generic one-loop amplitude
\eqn\eIgrav{{\cal
M}_{n}\sim\int d^{D}\ell \,{ \prod_{j=1}^{n-4} \ell\cdot q_{j}\over
\prod_{i=1}^n (\ell-k_{1\cdots i})^2} }
This suggests that a one-loop $n$-graviton ${\cal N}=8$
supergravity amplitude can be reduced to a sum of massive box
integrals multiplied by an operator of mass dimension eight and
$n$-gons, {\it i.e.,} ($n\geq 5$) scalar integrals evaluated in
dimensions $D+2k$
with $0\leq k \leq n-4$~\refs{\BernSV,\BjerrumBohrYW,\NastiSR,
\BernXJ} for $D>4$. In $D=4$ the no-triangle hypothesis suggest
that one-loop $n$-graviton amplitudes in $\cN=8$ do not contain
integral functions more singular than (massive) boxes and in
particular do not contain triangles nor bubble functions.
Including
the unexpected cancellations one will have in the generic case for
an arbitrary supersymmetric theory~\refs{\BernXJ}
\eqn\eNmax{\nu\leq 2n - (n-4+{\cal N})=
n+4-{\cal N}}
$\nu$ loop momenta in the numerator for theories with $0\leq
\cN\leq 8$ supersymmetries (replace $\cN$ by  ${\cal N}+1$
for an odd number of supersymmetries in~\eNmax).

The no-triangle hypothesis does carry cancellations  into  multi-loop
amplitudes. This can be
observed through cuts of amplitudes and through physical factorisation
limits linking $m<n$ loop amplitudes to $n$-loop amplitudes~\BernKD.
At multi-loop level  it has  been verified  that ${\cal  N} =  8$
supergravity is a finite theory  until three loops~\Bernhh.

In deriving the field theory limit of the $n$-gravitons amplitude at
genus one in string theory from the contributions of colliding
vertex operators~\refs{\MinahanHA,\BernPK} it is observed that the
total contribution to the $n$-graviton amplitude at one-loop is
composed of {\sl one-particle irreducible} contributions and
{\sl one-particle reducible}
contributions, see figure~1. These contributions
originates from the boundary of the moduli of the punctured Riemann
surface on which the string amplitude is
defined~\refs{\DHokerTA}.
\vskip2pt
$$
\includegraphics[width=12cm]{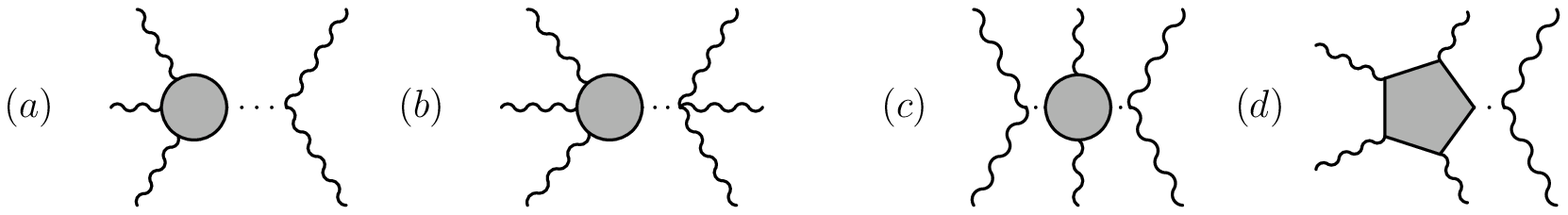}
$$
\begingroup
\narrower
\item{\bf\figred}\sl
Contribution from one-particle reducible graphs.
Up to and including five-graviton amplitudes the
reducible graphs are only
constructed from boxes, but
for six-graviton and beyond higher point amplitudes can occur
in the reducible part of the amplitude.
\endgroup
\vskip10pt

The one-particle reducible contributions, displayed in figure~1(b)
arise from the possibility of constructing a one-loop amplitude by
attaching $k$-point tree vertices to a one-loop $n-k$-point
amplitude in order to construct $n$-point contributions.

It  is a  well known  fact  that linearised  $\cN=8$ supersymmetry  of
perturbative  string  theory  guaranties  that  the  one-,  two-,  and
three-point amplitudes are vanishing at one-loop. Therefore in $\cN=8$
supergravity (and  type II  superstring theory) there  is no  room for
constructing reducible graphs from triangles or bubbles. This fact was
 noticed in~\KalloshYM.
This however does not imply the absence of triangles in
maximal supergravity gravity amplitudes because
supersymmetry allows higher-point reducible and irreducible amplitudes contributions that contain triangles.
 At one-loop order, the supersymmetric
cancellations enforced by the saturation of the sixteen fermionic
zero modes only subtract eight powers of loop momenta leading
to contributions of the type (again we display only the
contributions with the most powers of loop momentum)
\eqn\eMnn{ {\cal M}_{n}^{1PI}\sim {\cal O}_{8}\, \int d^D\ell {
\prod_{j=1}^{2(n-4)} \ell\cdot q_{j}\over \prod_{i=1}^n
(\ell-k_{1\cdots i})^2} }
where ${\cal O}_8$ is a mass dimension eight operator factorising in front
of the loop amplitude.
Hence we obtain triangle contributions after $(n-3)$ steps of
Passarino-Veltman reductions. No known explanation for cancellations
of triangles has been attributed solely to supersymmetry.

In  this  paper we will consider   the  explicit  computation   of  the
five-graviton  amplitude  at one-loop  in  maximal supergravity.   The
five-graviton one-loop  MHV amplitude  in four dimensions  has already
been derived  using the on-shell unitarity  methods in \refs{\BernML}.
The  method  used  in  the  present  paper is  different  and  is  not
restricted  to a particular  dimension. A  direct comparison  with the
results of that paper will appear in \refs{\BohrVanhove}.  We will use
the form  of the $n$-graviton  amplitude provided by the  field theory
limit  of  the  genus-one  amplitude compactified on a torus.  String  theory  allows  us  to
implement in a simple way  the effects of the $\cN=8$ supersymmetry by
using  the Jacobi identity  (and its  generalisation for  higher-point
amplitudes).   This provides  a practical  set-up for  classifying the
reducible contributions  and enables  us to recursively  construct the
$n$-point amplitude.  We discuss  the contributions that will occur in
higher-point amplitudes.

\newsec{The five-graviton amplitude}
\seclab\secfive
We will in this section consider the derivation of the five-point
amplitude in maximal supergravity in $D$ dimensions from the
field theory limit of type~II string theory compactified on
a $10-d$ dimensional torus.
In order to derive the five-graviton one-loop amplitude we
will use the rules of perturbative string theory at genus one.
A basic presentation of the employed string theory rules and
a discussion of the field theory limit are offered
in the appendices.
Further details will appear in~\refs{\BohrVanhove}.
The resulting field theory amplitude is given by an
irreducible contribution and a reducible contribution
displayed below in~\figfive(a) and~\figfive(b) respectively.
We will analyse these contributions in turn.
\vskip2pt
$$
\includegraphics[height=2cm]{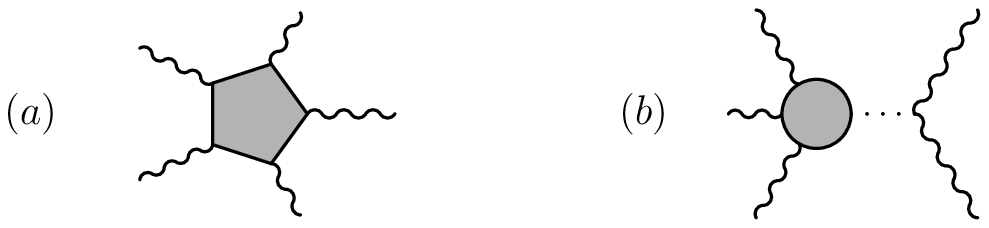}
$$
\begingroup
\narrower
\item{\bf\figfive}\sl Contribution to the field theory limit of the five-graviton amplitude.
\endgroup
\medskip

\subsec{The one-particle irreducible contributions}

The one-particle irreducible contribution to the five-graviton
amplitude at one-loop receives contributions both from the
$even/even$ spin structure sector of the genus-one string
theory amplitude and from the $odd/odd$ spin structure sector of the
amplitude.

Working out the bosonic contractions, the integrand of the five-point
amplitude in the $even/even$ spin structure sector takes the
form
\eqn\eFivep{\eqalign{
&A_5^{e/e}\!=\!{\kappa_{(10)}^2\over\ap^{{D\over2}-5}}\!\int_{{\cal
F}}{d^2\tau\over\tau_{2}}\, \Gamma_{(10-D,10-D)}\,
\prod_{i=1}^4\int_{{\cal T}} {d^2z_i\over\tau_2}\,\prod_{1\leq i<j\leq
5}\,|\chi(z_i-z_j)|^{-\ap\,k_{i}\cdot k_{j}}\times\cr
&\bigg(\bigg|T_{10}\cdot F^5\!\!+\!i\pi\!\! \sum_{i,j=1}^5\!\! k_i\cdot k_j
\partial_{i}\!\ln\chi(z_i-z_j)(t_8\cdot F^4_{\hat\imath})\bigg|^2
\!\!-\!\sum_{i\neq j} \! {k_{i}\cdot k_{j}\over\ap}
\partial_{i}\bar\partial_{j}\ln\chi(z_i-z_j)\,
(t_8\cdot F^4_{\hat\imath})(t_8\cdot F^4_{\hat\jmath})\bigg)  }}
The integrations in the above formula are over the positions of
the external states $z_i=\nu^{(1)}_i+i\tau_2\nu_i$ where the
domain of integration is ${\cal T}=\{|  \nu^{(1)}|\leq 1/2, \,
\nu\in [0,1]\}$ and $z_5=\tau$ by conformal invariance.
The factor $\Gamma_{(10-D,10-D)}$ represents the
contributions from the winding modes and Kaluza-Klein states.
Here $T_{10}\cdot F^5$ is the contribution from the contraction of
the ten world-sheet fermions defined in eq.~(A.9).
The quantity $t_8\cdot F^4_{\hat\imath}$ is defined as the
contractions of the field strengths $F_{\mu\nu}=h_{\mu}k_{\nu}-
h_{\nu}k_{\mu}$ (of the four states different from state $i$)
with the usual $t_8^{\mu_1\nu_1\cdots \mu_4\nu_4}$ tensor
defined in appendix~9.A of \refs{\GreenSP}.
The bosonic propagators are given by $\ln\chi(z)$ defined
in eq.~(A.6) of the appendix.
We refer to the appendix for our conventions and
for a further discussion of the field theory limit.

\medskip

The 1PI contributions to the field theory limit of the
amplitude will be obtained in the limit of $\ap\to0$ and
$\tau_2\to\infty$ while keeping $t=\ap\,\tau_2$ and
the distance between the vertex operators finite.
In this limit the fermionic and bosonic propagators are
\eqn\ePropinf{\eqalign{ S_{1}(z)&\ \to\
G_F(\nu)\ =\ \pi\,\sg(\nu)\cr
\partial_z\ln\chi(z) &\ \to\  \dot G_B(\nu)\ =\ \pi\nu- {1\over2}G_F(\nu)\cr
\partial_{z}\bar\partial_{z}\ln\chi(z)&\ \to\  -\ap\,{\pi\over4} \,{1\over t}}}
Taking $R\to0$ and $\ap/R\to0$ the lattice sum has the limit
\eqn\eGG{
\Gamma_{(10-D,10-D)}\ \to\ R^{5-{D\over2}}\, \tau_2^{5-{D\over2}}
}
We introduce the $n$-point integrals
\eqn\eNgon{
I_n^{(D)}[f(\nu)]\ \equiv\ \pi^{{D\over2}-n}\,\Gamma\left(n-{D\over2}\right)\, \prod_{i=1}^{n}\,
\int_0^1 d\nu_i\, f(\nu_i)\,Q_n(k_i)^{{D\over2}-n}\, \delta(\nu_n-1)
}
where
\eqn\eQn{ Q_n(k_i)\ = \sum_{1\leq i<j\leq n} \, (k_i\cdot k_j) \,
\Big[(\nu_i-\nu_j)^2-|\nu_i-\nu_j|\Big] }
The 1PI contribution to the five-graviton genus-one amplitude
 leads to the result
\eqn\eEEft{
{\cal M}_{5}^{1PI}\ =\
I_5^{(D)}\Big[|{\cal A}^{(1)\infty}_{5}|^2\Big]
+\pi\,I_5^{(D+2)}\Big[{\cal A}^{(2)\infty}_{5}\Big] }
This expression assumes the summation of all the ordering of the
external legs.
We refer to appendix~A.2 for further details. The second term
has a dimension shift from $D$ to $D+2$  which arises from the extra
inverse power of the loop proper time from the zero
mode contribution of the bosonic coordinates.\hfill\break

The various pieces of the field theory amplitudes are given by
\eqn\eFTPI{\eqalign{ {\cal A}^{(1)\infty}_{5}&\ =\
 t_{10}\cdot F^5+\pi\, \sum_{i\neq j} (h_{i}\cdot k_{j})
\dot G_B(\nu_i-\nu_j) \,(t_8\cdot F^4_{\hat\imath})\cr
&\ =\ t_{10}\cdot F^5-{\pi\over2}\sum_{i\neq j} (h_i\cdot k_j) \,
G_F(\nu_i-\nu_j)\,(t_8\cdot F^4_{\hat\imath}) -\pi \,H\cdot
K_{[5]}\cr
{\cal A}^{(2)\infty}_{5}&\ =\ \sum_{i\neq j}h_{i}\cdot \bar h_{j}\,
(t_8\cdot F^4_{\hat\imath})(t_8\cdot F^4_{\hat\jmath})\,
}}
Here $H$ and $\bar H$  defined as
\eqn\eH{
H\ =\ \sum_{i=1}^5 \, h_i\,(t_8\cdot F^4_{\hat\imath})\,,
\qquad \bar H\ =\ \sum_{i=1}^5 \bar h_i\,(t_8\cdot\bar F^4_{\hat
\imath}) \,
}
have been introduced together with
\eqn\eK{
K_{[n]}\ =\ \sum_{i=1}^n \,k_i\nu_i }
The quantity $t_{10}\cdot F^5$ (defined in eq.~(A.18)) depends on the
ordering of  the positions of the  vertex operators. It is defined as
the   field  theory   limit  of the  contractions between   the  fermions
$T_{10}\cdot F^5$ in the string theory amplitude.

From the contribution ${\cal A}_{5}^{(1)\infty}$ one gets
a combination of scalar pentagons $M_{5}[1]$, a combination
of linear pentagons $M_{5}[\nu]$ and a linear combination of
quadratic pentagons $M_{5}[\nu^2]$
\eqn\ePs{\quad
\hskip-0pt M_{5}[1]\ \ \ =\ I_5^{(D)}\Bigg[
\Big|t_{10}\cdot F^5
-{\pi\over2}\, \sum_{i\neq j} (h_{i}\cdot k_j) G_F(\nu_i-\nu_j) \,
(t_8\cdot F^4_{\hat\imath})\Big|^2\Bigg]
}
\vskip-17pt
\eqn\ePl{\quad \hskip-0pt
 M_{5}[\nu]\ \ \ \, =\ -\pi\,I_5^{(D)}\Bigg[
 \bigg(\!t_{10}\cdot F^5 -{\pi\over2}
\!\sum_{i\neq j} (k_{i}\cdot h_{j}) G_F(\nu_i-\nu_j) \,
(t_8\cdot F^4_{\hat\imath})\bigg) (H\cdot K_{[5]})\Bigg]
}
\eqn\ePq{\quad \hskip-195pt M_{5}[\nu^2]\ \, \, =\ \pi^2
\,\,I_5^{(D)}\Big[ \,\left(H\cdot K_{[5]}\right)(\bar H\cdot
K_{[5]}) \hfill\Big] }
\break
The expressions given here are
summed over all the ordering of the external legs. For
evaluating these expressions and extracting the various
contributions having branch cuts in different kinematic channels, one has to split the integral with respect to the
various orderings; see~\refs{\GreenSW,\gvstringloop,\GreenUJ}
and the appendix.

The contribution from ${\cal A}_{5}^{(2)\infty}$ in
eq.~\eEEft\ has an extra
power of $Q_{5}$ from the zero mode contribution of the bosonic
coordinates and contributes to a
linear combination of scalar pentagons in $D+2$
dimensions
\eqn\ePdS{\eqalign{
M^{(D+2)}_{5}[1]& \ =\  \sum_{i\neq
j}\, (h_{i}\cdot \bar h_{j})\,
 (t_8\cdot F^4_{\hat\imath}) \,(t_8\cdot F^4_{\hat\jmath})
\,I_5^{(D+2)}[1]
}}
The $odd/odd$ spin structure contribution to a toroidal compactification
of the $n$-graviton amplitude at one-loop vanishes in $D<10$. This is
because of the impossibility of saturating the fermionic zero modes
along the compactified directions with only external states
without polarisations along the internal directions.
The $odd/odd$ spin structure contributes to the amplitude
in ten dimensions for $n\geq5$ graviton amplitudes.
Its contribution have the following form \refs{\PeetersUB}
\eqn\eOOFt{ {\cal
M}_{5}^{o/o}\ \propto\  \epsilon^{\lambda\mu_1\cdots \mu_{9}}
\epsilon^{\lambda\nu_1\cdots\nu_{9}}
\prod_{1\leq r\leq 5} h^{(r)}_{\mu_r} \bar h^{(r)}_{\nu_r}\,
\prod_{1\leq s\leq 4} k^{(s)}_{\mu_{5+s}} k^{(s)}_{\nu_{5+s}}
\, I_5^{(12)}[1]
}

\medskip
\noindent Only the quadratic pentagons in eq.~\ePq\ can contain
triangles, and we will show in section~4 how of these contributions
cancels explicitly.

\subsec{The reducible contribution }

We now  turn to the 1PR  contributions. We shall see  that they cannot
contribute to  triangles at  this order since  they are only  given by
massive scalar box contributions represented in~\figfive(b).

The reducible expressions arise when two (or more) vertex operators
collide in the field theory limit.
When $z_{i}\to z_{j}$ the bosonic or fermionic propagator develop a pole
\eqn\ePole{ \lim_{i\to j}
\partial_{i}\ln\chi(z_{i}-z_{j})\ =\ -{1\over4}\, {1\over z_i-z_j},\qquad
 \lim_{i\to j}
S_{1}(\bar z_i-\bar z_j)\ =\ {1\over \bar z_i-\bar z_j} }
and the integrand  $|{\cal A}^{(1)}_{5}|^2$ in eq.~\eFivep\
can develop a pole when, say $z_4\to z_5$
\eqn\eGfive{\eqalign{ \lim_{4\to5} |{\cal
A}^{(1)}_{5}|^2&\ \sim\  {1\over |z_{45}|^2}\, |t_{(45)}|^2 }}
where
\eqn\eTTT{
 t_{(45)}\ \equiv\  t_{8}(F^1 F^2 F^3
F^{45})-h_{5}\cdot
k_{4} \,(t_8\cdot F^4_{\hat 5})+h_{4}\cdot k_{5} \,(t_8\cdot F_{ \hat 4}^4)
}
with $F^{45}_{\mu\nu}= (F^4)_{\mu}{}^\lambda \, (F^5)_{\lambda\nu}$.
The sum of the reducible contributions is given by

\eqn\eAevenTwo{\eqalign{ A_{5,1PR}^{e/e}&\ =\ {\kappa_{(10)}^2\over\ap^{{D\over2}
-5}}\sum_{i\neq j}\int_{\cal F}
{d^2\tau\over\tau_{2}}\,\Gamma_{(10-D,10-D)}\cr
&
 \prod_{r=1}^3\int_{\cal T}{d^2z_{r}\over\tau_2}\,
  { |t_{(ij)}|^2\over |z_i-z_j|^{2+2\alpha' k_{i}\cdot k_{j}}}
\, \prod_{1\leq u<v\leq 4}\, |\chi(z_u-z_v)|^{-2\ap P_{v}\cdot
P_{u}} }}
with $z_4=\tau$ and $P_m=k_m$ if
$m\neq i$ or $m\neq j$ and $P_i=k_i+k_j$.
The  integration over
$z_{i}=z_{j}+\zeta$ with $|\zeta|<\epsilon\ll1$
gives  in the field theory limit \refs{\MinahanHA, \BernPK}
\eqn\eSinglePole{\eqalign{ \lim_{\ap\to0}\int_{|\zeta|<\epsilon}
d^2\zeta |\zeta|^{-2\alpha' k_{i}\cdot k_{j}-2} &\ =\  -
\lim_{\ap\to0}{\epsilon^{\ap k_{i}\cdot k_{j}}\over \ap\, k_{i}\cdot
k_{j}}\  =\ -{1\over \ap\, k_{i}\cdot k_{j}} }}
Therefore the reducible contribution to the five-point amplitude is
given by
\eqn\eAevenTT{ {\cal
M}_{5}^{1PR}\ =\ \lim_{\ap\to0}\kappa_{(D)}^{-2}{\cal A}^{e/e}_{5,1PR}=
\pi^{D-8\over2}\Gamma\left(8-D\over2\right)\,\sum_{i\neq j} t_{(ij)}\,
\prod_{r=1}^4\int_{0}^1{d\nu_{r}}\,
Q_{4}(P_i)^{{D\over2}-4}\,\delta(\nu_4-1) }
which is the sum of contributions from one-mass scalar boxes
 evaluated for the external momenta $(P_{1},P_{2},P_{3},P_{4})$.

\newsec{Reduction formulas}

In this section we will discuss the integral reduction
formulas needed to examine the integral contributions
of the five-point amplitude.
The reduction formulas presented in~\refs{\BernEM,\BernKR}
could have been used in this analysis, however we found
it useful to derive the reduction formulas (which
have their root in gauge invariance and
the decoupling of longitudinal modes) from the viewpoint of
string theory.

We will consider special expressions at genus one between very
specific vertex operators. The vertex operators are not describing
physical external states but are part of the physical vertex
operators of string theory. The identity we will derive in
this section will be entering the analysis of the graviton
five-point amplitude.

We introduce the fermionic vertex operators (see the appendix for
definitions and conventions)
\eqn\eVPsiPsi{
V_{\psi\bar\psi}^{IJ,KL}(k)\ =\  \int_{{\cal T}} d^2z\,
\ :\psi^{I}\psi^{J}\,\bar\psi^{K}\bar\psi^{L} \, e^{ik\cdot x(z)}:
}
the mixed fermionic and bosonic vertex operators with a longitudinal part

\eqn\eVdXPsi{\eqalign{
V_{\partial x\bar\psi}^{KL}(k)&\ =\ \int_{{\cal T}} d^2z\
:\partial\big(\bar\psi^{K}\bar\psi^{L} \,e^{ik\cdot x(z)}\big)
: \ \ =\  \int_{{\cal T}} d^2z\ :ik\cdot\partial x\,\bar\psi^{K}\bar\psi^{L}
\,e^{ik\cdot x(z)}:\cr
 V_{\psi\bar\partial x}^{IJ}(k)&\ =\ \int_{{\cal
T}} d^2z\ :\bar\partial\big(\psi^{I}\psi^{J} \,e^{ik\cdot x(z)}\big): \ \
=\ \int_{{\cal T}} d^2z\ :ik\cdot\bar\partial X\,\psi^{I}\psi^{J}
\,e^{ik\cdot x(z)}:  }}
These expressions are vanishing because the left-moving or the right-moving
part of these vertex operators
is purely longitudinal and the torus
${\cal T}=\{z=\nu^{(1)}+i\tau_2\nu ; |\nu^{(1)}|\leq 1/2; \nu\in [0,1]\}$
has no boundaries.
We introduce as well  the purely longitudinal bosonic vertex operator
\eqn\eVdXdX{\eqalign{ V_{\partial X\bar\partial X}(k)&\ =\ \int_{{\cal
T}} d^2z\ :\partial\bar\partial\big(e^{ik\cdot x(z)}\big):\ \ =\ -\int_{{\cal T}}
d^2z\ :k\cdot\partial x\,k\cdot\bar\partial x\,e^{ik\cdot x(z)}:
}}
In these expressions $x^\mu(z)$  and $\psi^\mu(z)$ are the
conformal fields of weight 0 and $1/2$ of  the RNS formulation of
perturbative string theory. The index $\mu$ runs
from 0 to $D\leq10$. The manipulations in this section will be done
using the rule for computing correlators at genus-one order in
string theory, but the manipulations here do not require that we are
working in the critical dimension $D=10$ neither that we are working
with physical vertex operators. In this section
the lattice factor $\Gamma_{(10-D,10-D)}$ has
been replaced by its field theory approximation
$\tau_2^{5-D/2}$ of eq.~\eGG. This scheme was already used in the
so-called `string based rules' of~\refs{\StringBased}.

We introduce the following notation
\eqn\ePsieight{ {\cal
O}_{\psi^{2n},\bar\psi^{2n}}(k_{i_1},\dots,k_{i_n})\ =\
t_{I_{i_1}\cdots I_{i_n}}^L t_{J_{i_1}\cdots J_{i_n}}^R \,
V_{\psi\bar\psi}^{I_{i_1}I_{i_2},J_{i_{1}}J_{i_{2}}}(k_{i_1})\cdots
V_{\psi\bar\psi}^{I_{i_{n-1}}I_{i_n},J_{i_{n-1}}J_{i_n}}(k_{i_n})
}
where $t^L_{I_1\cdots I_n}$ and $t^R_{I_1\cdots I_n}$ are rank
$n$-tensors contracting the Lorentz indices of the left moving
fermions $\psi^\mu(z)$ and the right moving fermions
$\bar\psi^\mu(z)$.

We will start by considering the genus one expression involving
four fermionic operators evaluated in the $even/even$ spin structure
sector.
The result is
\eqn\eAI{ \left\langle {\cal
O}_{\psi^{8},\bar\psi^{8}}(k_{1},\dots,k_{4})\right\rangle_{e/e}\ =\ t_{8}t^L\,
t_{8}t^R\, \int_{{\cal
F}}{d^2\tau\over\tau_{2}}\,\tau_{2}^{4-{D\over2}}
\,\prod_{i=1}^3\int_{\cal T} {d^2 z_i\over\tau_2}\, \prod_{1\leq
i<j\leq 4}\, |\chi(z_i-z_j)|^{-\ap\,k_{i}\cdot k_{j}} }
with $z_4=1$. This result is proportional to the genus-one four-point amplitude
in type II superstring which has
the field theory limit $\ap\to0$ in $D=d-2\epsilon$ dimensions. The
one-loop four point scalar box $I_{4}^{(D)}[1]$ (in the dimensional
regularisation scheme) is summed over all the possible ordering of
the external legs \refs{\GreenSW}

\eqn\eAIft{ \lim_{\ap\to0}\left\langle {\cal
O}_{\psi^{8},\bar\psi^{8}}(k_{1},\dots,k_{4})\right\rangle_{e/e}\ = \
t_{8}t^L\, t_{8}t^R\,I_4^{(D)}[1] }
where $ I_4^{(D)}[1]$ is defined in eq.~\eNgon\ and, {\it e.g.,} $t_8 t^L$
is the contraction of $t_8$ and $t^L$.
This expression is the sum of the
$s$-channel $I_4(s,t)$, $t$-channel $I_4(t,u)$ and $u$-channel
$I_4(u,s)$ boxes \refs{\GreenSW,\gvstringloop,\GreenUJ}.

\medskip
We consider now the $even/even$ spin structure correlator with the
insertion of two longitudinal vertex operators $h^5_{ij} \,
V^{ij}_{\partial X\bar\psi}=0$
and $h^4_{ij} V^{ij}_{\psi\bar\partial X}=0$ which we defined
in eq.~\eVdXPsi. Now
\eqn\eAIII{\eqalign{
0&\ =\ h^5_{ij}h^4_{kl}\left\langle V_{\partial X \bar\psi}^{ij}(k_{5})
V_{\psi\bar\partial X}^{kl}(k_{4}) {\cal
O}_{\psi^{6},\bar\psi^{6}}(k_{1},\dots,k_{3})
\right\rangle_{e/e}\cr
&\ =\ \int_{\cal F}{d^2\tau\over\tau_2^2}\, \tau_2^{5-{D\over2}}\,
\prod_{i=1}^4\int_{\cal T}{d^2z_i\over\tau_2}\, \prod_{1\leq i<j\leq
5}t_8(h^5\, t^L)\,t_8(h^4 \,t^R)\, |\chi(z_i-z_j)|^{-\ap\,k_i\cdot
k_j}\times\cr & \bigg[\!\bigg(\sum_{j=1}^5\, ik_5\cdot k_j\,
\partial_{5}\ln\chi(z_5-z_j)\bigg) \!\bigg(\sum_{i=1}^5\, ik_4\cdot k_i\,
\bar\partial_{4}\ln\chi(z_4-z_i)\bigg) -{k_{5}\cdot k_{4}\over\ap}
\partial_{5}\bar\partial_{4}\ln\chi(z_5-z_4) \bigg]
}}
The contractions of the eight left-moving and eight right-moving
fermions and the sum over the spin structure have been done using
the Jacobi identity given in the appendix.  It is important
to notice that this gives a contribution that is a constant
independent of the positions of the
vertex operators. Thus eq.~\eAIII\ implies that
\eqn\eZero{\eqalign{ 0&\ =\ {\cal R}\ \equiv\ \int_{{\cal
F}}{d^2\tau\over\tau_{2}}\,\tau_{2}^{5-{D\over2}} \,
\prod_{i=1}^4\int_{{\cal T}} d^2\nu_i\,\prod_{1\leq i<j\leq
5}|\chi(z_i-z_j)|^{-\ap\,k_{i}\cdot k_{j}}\times\cr
&\bigg[\!\bigg(\sum_{j=1}^5\, ik_5\cdot k_j\,
\partial_{5}\ln\chi(z_5-z_j)\bigg) \!\bigg(\sum_{i=1}^5\, ik_4\cdot k_i\,
\bar\partial_{4}\ln\chi(z_4-z_i)\bigg) -{k_{4}\cdot k_{5}\over\ap}
\partial_{5}\bar\partial_{4}\ln\chi(z_5-z_4) \bigg]  }}
In the field theory limit this amplitude gives rise to the
one-particle irreducible (1PI) contributions and one-particle
reducible contributions (1PR).

The 1PI contribution is obtained using the field theory asymptotic
of the bosonic and fermionic propagators given in eq.~\ePropinf.
With the same manipulation as for the 1PI contribution to the
physical amplitude in eq.~\eFivep\ we obtain
\eqn\eOPI{\eqalign{
{\cal R}^{1PI}&\ =\ -\, I_5^{(D)}\Big[(k_4\cdot K_{[5]})(k_5\cdot
K_{[5]})\Big]\cr 
&\ -{1\over2}\ I_5^{(D)}\Big[(\sum_{i=1}^5\,(k_5\cdot k_i)\, \sg(\nu_5-\nu_i))
(k_4\cdot K_{[5]})\Big]+
(4\leftrightarrow 5)\cr
&\  -{1\over4}\,I_5^{(D)}\Big[\sum_{i,j=1}^5\,(k_5\cdot  k_i)\,\sg(\nu_5-\nu_i)
(k_4\cdot  k_j)\,\sg(\nu_4-\nu_j)\Big]\cr
&\, -\,(k_4\cdot k_5)\, I_5^{(D+2)}[1] }}
 The expression in eq.~\eOPI\ is the sum of scalar,
linear and quadratic pentagons in dimension $D$ and a scalar pentagon
in dimension $D+2$ from the zero mode contribution from the
correlator between  the bosonic  coordinates. 

\medskip

The reducible contributions (1PR) in the field theory limit
of ${\cal R}$ are obtained only when the bosonic propagators
develop a pole as in eq.~\ePole\ when $z_{4}\to z_{5}$.
Note that when $z_{4}\to z_m$ or $z_5\to z_m$ with $m=1,2,3$
the expression~\eZero\ behaves as $1/(\bar z_4-\bar z_m)$ and $1/(z_5-z_m)$
respectively, which does not lead to a reducible contribution
because this requires a $1/|z_i-z_j|^2$ type of singularity as described in eq.~\eSinglePole.
In this case the expression ${\cal R}$ behaves as
\eqn\eAevenTwo{\eqalign{
{\cal R}^{1PR}&\ =\ \int_{\cal F}
{d^2\tau\over\tau_{2}}\,\tau_{2}^{5-{D\over2}}
 \prod_{r=1}^4\int_{\cal T}{d^2z_{r}\over\tau_2}\,
  { (k_4\cdot k_5)^2\over |z_i-z_j|^{2+2\alpha' k_{i}\cdot k_{j}}}
\, \prod_{1\leq u<v\leq 4}\, |\chi(z_u-z_v)|^{-2\ap P_{v}\cdot
P_{u}} }}
with $\{P_m\}=\{k_1,k_2,k_3,k_4+k_5\}$. Performing the integration over
$z_{5}=z_{4}+\zeta$ with $|\zeta|<\epsilon\ll1$ as in eq.~\eSinglePole\
the 1PR contribution to ${\cal R}$ is given by the one-mass scalar
box obtained by colliding the states 4 and 5
\eqn\eOPR{ {\cal R}^{1PR}\ \equiv\ \lim_{\ap\to0}\lim_{4\to 5}{\cal R}\ =\
-(k_4\cdot k_5)\, I_4^{(45)}[1] }

\medskip

Collecting the 1PI and 1PR contributions to the field theory limit
of~\eZero\ gives the following identity
\eqn\eTotal{\eqalign{
I_5^{(D)}[(k_4\cdot K_{[5]})(k_5\cdot K_{[5]})]&\ =\ (k_4\cdot k_5)\,
 I_5^{(D+2)}[1]\cr
&\  +{1\over2}\,I_5^{(D)}\Big[(\sum_{i=1}^5\,(k_5\cdot  k_i)\,\sg(\nu_5-\nu_i))
(k_4\cdot K_{[5]})\Big]+(4\leftrightarrow 5)\cr
&\  +{1\over4}\,I_5^{(D)}\Big[\sum_{i,j=1}^5\,(k_5\cdot  k_i)\,\sg(\nu_5-\nu_i)
(k_4\cdot  k_j)\,\sg(\nu_4-\nu_j)\Big]\cr
 &\ +(k_4\cdot
k_5)\, I_4^{(45)}[1] }}
relating a linear combination of quadratic pentagons to a scalar
pentagon in $D+2$ dimensions and scalar and linear pentagons
as well as one-mass boxes in $D$ dimensions.
The loop integral is defined with the summation over all the orderings
and the right-hand-side does not contain any triangles.
The same identity is valid for any choice of a pair of
momenta $k_m$ and $k_n$ with $(m,n)\in\{1,2,3,4,5\}^2$. In~\eTotal\
we had $k_m=4$ and $k_n=5$.

Similar relations as~\eTotal\ were found in section~6
of~\refs{\BernKR} using manipulations of Feynman
parameter integrals with a fixed ordering of the external legs.
Via further reduction of linear pentagons to one-mass boxes
it can be observed that $D+2$ pentagons are not present
in the amplitudes in $D = 4$~\refs{\BernKR}.

\newsec{Cancellation of the triangles}
\seclab\secnotriangle

We will now show how the identity given in eq.~\eTotal\ allows
us to remove the potential triangle contributions present in the
quadratic pentagon $M_5[\nu^2]$ in eq.~\ePq\ for the five-graviton amplitude in $\cN=8$ supergravity.

For an amplitude with at least five external states there are at
least four independent momenta, say $k_{1}$, $k_{2}$, $k_{3}$ and
$k_{4}$, in dimension $D\geq 4$.
Hence we can decompose $H$ and $\bar H$
in such a basis as
\eqn\eHdec{ H \ =\ \sum_{i=1}^4 c_{i }\,
k_{i}+q_{\perp},  \qquad \bar H \ =\ \sum_{i=1}^4 \bar c_{i }\,
k_{i}+\bar q_{\perp}}
where $c_i$ and $\bar c_i$ are constants and $q_{\perp}$ and
$\bar q_{\perp}$ are orthogonal to the chosen four independent
momenta of the external states (this is needed only in $D>
4$). We have assumed a generic configuration of
external momenta with no momenta being collinear.
The case of collinear momenta is correctly captured by the
reduction formulas.

Plugging this decomposition into eq.~\ePq\ the combination
of quadratic pentagons can be rewritten as the linear
combination

\eqn\ePqBis{ M_{5}[\nu^2]\ \propto\  \sum_{i,j=1}^4 c_{i}\bar c_{j}\,
I^{(D)}_{5}\Big[(k_{i}\cdot K_{[5]})\, (k_{j}\cdot K_{[5]})\Big] }
of the same quantities appearing in the left-hand-side of the identity
in eq.~\eTotal.
Because the right-hand-side of this identity does not have any
triangle contributions, we conclude that the five-graviton amplitude
$M_5^{1PI}$ of eq.~\eEEft\ does not contain any triangles.

It should be noted that we have used the reduction formula given
in the form of eq.~\eTotal\ directly without having to solve for
individual quadratic pentagons. We also note that we did not have
to invert the Gram determinant of the external momenta which is
very messy at higher-point order because of the linear
dependence in the kinematic invariants~\refs{\BernEM,\BernKR}.

\newsec{Cancellation of triangles in higher-point amplitudes}

At six-point order the integrand of the amplitude takes the
recursive form (see~\refs{\BohrVanhove} and the appendix)
\eqn\eAintSix{ {\cal A}_{6}\ =\  T_{12}\cdot F^6+ \sum_{i} (h_{i}\cdot
\partial X)\,
 (T_{10}\cdot F^5_{\hat\imath})
+\sum_{i\neq j} (h_{i}\cdot \partial X) \, (h_{j}\cdot \partial X)\,
(t_8\cdot F^4_{\hat\imath,\hat\jmath}) }
where the quantity $t_8\cdot F^4_{\hat\imath,\hat\jmath}$ is the
four point amplitude constructed from the field strengths of the
four external states different from $i$ and $j$.

The reducible graphs are given by the one-mass
box of~\figred(b) and the two-mass boxes of~\figred(c)
(depending on the ordering of the vertices around the loop this
gives the two-mass easy or hard scalar
box~\refs{\BernML,\BernSV,\BjerrumBohrYW,\NastiSR}) and the one-mass pentagon of~\figred(d).
Quadratic pentagons in the reducible part of the six-point
amplitude can appear from poles arising from the second and the third
term in~\eAintSix. In case of a pole from colliding the states $5$ and
$6$ we have the quadratic pentagons
\eqn\eSixPR{ {1\over
s_{56}}\,I_5^{[D]}\Big[(H^{(6\to5)}\cdot K^{(6\to5)}_{[6]}) (\bar
H^{(6\to5)}\cdot K^{(6\to5)}_{[6]})\Big] }
Here $K_{[6]}^{(6\to5)}$
is the five-point sum $K_{[5]}$ for the momenta
$\{k_1,k_2,k_3,k_4,k_5+k_6\}$, and
\eqn\eHH{H^{(6\to5)}\ =\ \sum_{i=1}^4\, h_i\, t_{(56)\hat\imath} }
is  a  linear  combination  of  the  polarisations  weighted  by  the
five-point tensor  $t_{(56)\hat\imath}$ defined as  in eq.~\eTTT\ for
the external states different  from $i$. Decomposing the tensor~\eHH\
as a linear combination  of $k_1$, $k_2$, $k_3$ and  $k_4$ the
analysis
of   sections~\secfive\ and~\secnotriangle\   assures   that   this
contribution has no triangles.

So in the six-graviton amplitude the triangle can only be present
in the irreducible part.
The total amplitude the six-point
amplitude contains  two types of contributions, depending on whether
there are contractions between left-moving $\partial x$ and
right-moving $\bar\partial x$ or not. The term involving the
left/right contraction are
\eqn\eAsixLR{\eqalign{ {\cal
A}_{6}^{(2)}\ & =\  \sum_{i\neq j} (h_{i}\cdot \bar h_{j})
{1\over\ap\tau_{2}}  (T_{10}\cdot F^5_{\hat\imath}) (T_{10}\cdot
F^5_{\hat\jmath})+\!\sum_{i\neq j\atop p\neq q}  (h_{i}\cdot \bar
h_{j}) (h_{p}\cdot \bar h_{q}) {1\over(\ap\tau_{2})^2}  (t_{8}\cdot
F^4_{\hat\imath,\hat\jmath})\, (t_{8}\cdot F^4_{\hat p,\hat q})\cr
&+\sum_{i\neq j\atop p,q}  (h_{i}\cdot \bar h_{j})\,(\bar h_{p}\cdot
k_{q})\,{1\over\ap\tau_{2}}\,\bar\partial\ln\chi(z_p-z_q)
 (T_{10}\cdot F^5_{\hat\imath})\, (t_{8}\cdot F^4_{\hat p,\hat \jmath}) +c.c.\cr
&+\!\!\!\sum_{i\neq j\atop p,q,m,n}\!\!\!\!(h_{i}\cdot \bar h_{j})\, (h_{p}\cdot k_{m})\,
(\bar h_{q}\cdot k_{n})\,{1\over\ap\tau_{2}}\,
\partial\ln\chi(z_p-z_m)\bar\partial\ln\chi(z_q-z_n)
(t_{8}\cdot F^4_{\hat p,\hat \imath})\, (t_{8}\cdot F^4_{\hat q,\hat \jmath}) \cr
}}
In the field theory limit this expression leads to 1PI contributions
composed by a sum of scalar,
linear and quadratic hexagons evaluated in dimension $D+2$ and a
scalar hexagon evaluated in dimension $D+4$.
None of these contributions contain triangles.

\medskip
The other 1PI contributions to the six-point amplitude can be written as
\eqn\eAsixOne{
{\cal A}_{6}^{(1)\infty}\ =\ \bigg|t_{12}\cdot F^6+i \sum_{i,m=1}^6
(h_{i}\cdot k_{m})\,
\dot G_B(\nu_i-\nu_m)\, {\cal A}_{5(\hat\imath)}^{(1)\infty}+
\sum_{i\neq j} (h_{i}\cdot h_{j})\,
{\pi\over\ap \tau_{2}}\,  (t_{8}\cdot
F^4_{\hat
\imath,\hat\jmath})\bigg|^2
}
where ${\cal A}_{5(\hat\imath)}^{(1)\infty}$ is the five-point
amplitude given in eq.~\eFTPI\ evaluated for the five external
states different from $i$

\eqn\eFTPIinot{
 {\cal A}^{(1)\infty}_{5(\hat\imath)}
\ =\ t_{10}\cdot F^5_{\hat\imath}-{\pi\over2}\sum_{i\neq j} (h_j\cdot k_m) \,
G_F(\nu_j-\nu_m)\,(t_8\cdot F^4_{\hat\imath\hat\jmath}) -\pi \,H_{\hat\imath}\cdot K_{[6]}
}
where $H_{\hat\imath}$ is defined as in eq.~\eH,
\eqn\eHH{
H_{\hat\imath}\ =\ \sum_{i\neq j} h_j \, t_8F^4_{\hat\imath\hat\jmath}
}
and  $K_{[6]}$ is the total momentum defined in eq.~\eK.
The only pieces that could lead to
triangles at six-point arise from the contributions
\eqn\eHcc{\eqalign{
\delta A_{6}^{(1)\infty}& \ = \ \sum_{i,m=1}^6 h_{i}\cdot k_{m}\, (\nu_i-\nu_m)\,
{\cal A}_{5(\hat\imath)}^{(1)\infty}\cr
&\ =\ -\bigg(\sum_{i=1}^6 \, h_i \big[t_{10}\cdot F^5_{\hat\imath}-{\pi\over2}\sum_{i\neq j} (h_j\cdot k_m) \,
G_F(\nu_j-\nu_m)\,(t_8\cdot F^4_{\hat\imath\hat\jmath})\big]\bigg)
\cdot K_{[6]}\cr
&\ +\pi \sum_{i=1}^6 (h_i\cdot K_{[6]})\, (H_{\hat\imath}\cdot K_{[6]})
}}
The sum over the polarisations can be decomposed on a basis of independent
momenta (as in eq.~\eHdec) as
\eqn\eHdecSix{
\sum_{i=1}^6 \, h_i t_i=
\cases{
\sum_{i=1}^4 \, c_i\, k_i & for $D=4$\cr
\sum_{i=1}^5\,c_i\, k_i+q_\perp& for $D\geq5$
}}
where the  coefficients $c_i$  are constants and  $t_i$ is  either the
combination   multiplying   $h_i$   in~\eHcc\   or
$H_{\hat\imath}$. The constants for each tensorial structure do not have to be
identical.   These   expressions   lead  to   cubic   hexagons
$I_6^{(D)}\big[(k_i\cdot   K_{[6]})(k_j\cdot  K_{[6]})(k_l\cdot   K_{[6]})\big]$  or
quartic  hexagons  $I_6^{(D)}\big[(k_i\cdot K_{[6]})(k_j\cdot
K_{[6]})(k_l\cdot
K_{[6]})(k_m\cdot K_{[6]})\big]$ that will have to be cancelled by implementing the
reductions formulas for the six-point integrals \refs{\BohrVanhove}.
%

\newsec{Discussion}

In this paper we have explored the one-loop $n$-graviton
amplitude derived in the field theory limit ($\alpha'\to0$)
of type IIA and IIB string theory while preserving maximal
supersymmetry of the theory.

In this `string based' formalism the integrand of the one-loop
amplitudes in supergravity takes the form of the square of
corresponding super-Yang-Mills amplitudes plus an additional
contribution from the zero modes of the bosonic coordinate coupling
the left and right moving sectors. We have shown that triangle 
 integral functions are not present at one-loop, in
accordance with the ``no-triangle hypothesis.''

The Kawai, Lewellen and Tye relations~\refs{\Kawaixq}, which are
derived from string theory, express gravity tree amplitudes as the
sums of products of two Yang-Mills tree amplitudes and have many exciting and
surprising consequences. Tree-level gravity amplitudes were
shown in~\refs{\BernBB} to enjoy enhanced symmetries similar to
those found for Yang-Mills theories inherited via the KLT relations.
The surprising good high-energy behaviour of gravity tree
amplitudes~\refs{\Treegravity,\BjerrumBohrXX,\BjerrumBohrJR} and the
cancellations of certain tree-graphs has been linked to the
cancellations of integral functions at one-loop level
in~\refs{\BjerrumBohrYW,\BernXJ}. The good high-energy behaviour of gravity
amplitudes at tree level was recently attributed to
basic gauge symmetry of the underlying gravitational
Lagrangian~\refs{\ArkaniHamedYF}. Gauge invariance was first linked
to unexpected cancellations for loop and tree amplitudes in gravity
theories in ref.~\refs{\BernXJ}.

In this paper we have investigated the cancellations of integral
functions at one-loop level encapsulated by the ``no-triangle
hypothesis'', from the viewpoint of the field theory limit of string
theory. We would like to emphasise at this point that
the viewpoint of using string theory as a guideline for
calculations in analysing the ``no-triangle hypothesis'' is
very different from that of unitarity methods. Our conclusions
however remain the same.
The origin of triangle cancellation has in this paper been
attributed to the decoupling of the longitudinal modes of string
theory in the field theory limit and the summation
 over all the possible orderings of the external legs due to the absence
of colour ordering in gravity amplitudes. In the language of field
theory this means that the ``no-triangle'' cancellations has their
roots in the gauge invariance of the theory as it appear to be the
case for the tree level amplitude simplifications. The cancellations
caused by the identities decoupling the longitudinal modes should
also apply to the pure spinor formulation of string perturbation
theory~\refs{\BerkovitsVC}.

At multi-loop level cancellations such as the ones observed
at one-loop level might have the potency to ultimately lead to
a ultra-violet finite point-like theory of perturbative gravity
in four dimensions as was suggested
in~\refs{\BjerrumBohrYW,\GreenGT,\BernKD,\GreenYU,\BernXJ}.
Cancellations of potential
UV-divergences at three-loop level was examined in~\refs{\Bernhh}
and by explicit computation it was shown at three-loops
that the UV-behaviour of maximal supergravity in $D=4$ is
no worse than that of $\cN=4$ super-Yang-Mills.

It is surprising that gauge invariance appears to be the main
driving force for the observed simplifications of gravity tree and
loop amplitudes. The full symmetry principle behind these unexpected
cancellations appear to have the potency to lead to new ground
breaking discoveries regarding the UV-behaviour of perturbative
gravity. Further investigations are clearly needed -- especially at
multi-loop level using the explicit information about the origin of
cancellations at tree and loop level.

\bigskip\bigskip
\noindent {\bf Acknowledgements:}
We would like to thank Zvi Bern and Lance Dixon for important
feedback on this paper. We would also like to thank Simon
Badger, Anirban Basu, Kasper Risager and Paolo di Vecchia
for useful discussions. Both authors would like to thank the
Niels Bohr Institute in Copenhagen for hospitality.
The author (PV) would like to acknowledge NORDITA for financial
support and would like to thank the organisers of the 2007 Les
Houches School on string theory for providing an inspiring
atmosphere where part of this work was taken out.
This research was supported in part (NEJBB) by the grant
DE-FG0290ER40542 of the US Department of Energy and   (PV) the
RTN contracts  MRTN-CT-2004-005104 and by the ANR grant BLAN06-3-137168.

\appendix{A}{The $n$-graviton amplitude at genus one in type II
string theory}

We compute the $n$-graviton amplitude at one loop in type IIA/B
string theory in ten dimensions. With the following normalisations
of the world-sheet action for the type II superstring we have
\eqn\eRNSaction{ S= {1\over 2\pi \alpha'}\, \int d^2z \, (\partial
x^\mu\bar\partial x_{\mu}+\psi^\mu \bar\partial \psi_{\mu}+\bar\psi^\mu
\partial \bar\psi_{\mu}) }
where $\alpha'=\ell_{s}^2$, and the graviton
vertex operator in the $(0,0)$-ghost picture is
\eqn\eV{
V^{(0,0)}={\kappa_{(10)}\over \alpha'}\, :h_{\mu\nu}\,
(\partial x^\mu +i \, k\cdot \psi \psi^\mu)
\,(\bar\partial x^\nu+i\, k\cdot \bar\psi \bar\psi^\nu)\, e^{i\,k\cdot x}:
}
We define $\kappa_{(10)}^2 = 2^6\pi^7 \, \ap^4$ \refs{\gvstringloop}.
The symmetric polarisation tensor $h_{\mu\nu}$ is decomposed as
$ (h_{\mu}\bar h_{\nu}+h_{\nu}\bar h_{\mu})/2$
where $h_\mu$ and $\bar h_\mu$ are
polarisation vectors satisfying the transversality
condition $k^\mu h_\mu=0$ and $k^\mu\bar h_\mu=0$.

The string theory $S$-matrix for an $n$-graviton amplitude is
expanded as
\eqn\eSm{ {\bf A} = \kappa_{(10)}^{n-2} \, g_s^n\,
\left({1\over g_s^2}\, A^{\rm tree}+ 2\pi A^{\rm genus-1}+\cdots
\right) }

\subsec{General structure of the amplitude}

The general structure of the multi-graviton one-loop amplitude in type
IIA/B string theory  compactified to $D$ dimensions on a $10-D$-torus
is given by the sum of the $even/even$ and the $odd/odd$ spin structure contribution
$A^{\rm genus-1}=A^{e/e}_{n}+A^{o/o}_{n}$.
The $even/even$ spin structure contribution takes the form~\refs{\DHokerTA}

\eqn\eOneeL{
A^{e/e}_{n}={\kappa^2_{(10)}\over \ap^{{D\over2}-n}}\,\int_{\cal F}{
d^2\tau\over \tau_{2}}\,  \tau_{2}^{n-5} \,\Gamma_{(10-D,10-D)}\,
 \prod_{i=1}^{n-1}\int_{\cal T} {d^2{z}_{i}\over \tau_{2}}\,
\left\langle|{\cal A}_{n}|^2\, \prod_{i=1}^n e^{i\, k\cdot x(z_{i})}
\right\rangle
}
where $z_i=\nu^{(1)}_i+i\tau_2\,\nu_i$
with $-1/2\leq \nu^{(1)}_i\leq 1/2$,
$0\leq \nu_i\leq 1$ and $z_n=\tau$. $\Gamma_{(10-D,10-D)}$ is
defined as the lattice sum over the
winding modes and Kaluza-Klein states of the
type~II string compactified on a $10-D$-torus.
The integrand takes the form \refs{\BohrVanhove}
\eqn\eAn{\eqalign{
{\cal A}_{n}&=  T_{2n}\cdot F^n+\sum_{i=1}^n h_{i}\cdot \partial
x(z_{i}) \,{\cal A}_{n-1}(\hat\imath)
}}
where the $\hat\imath$ denotes that state $i$ is not included. The
bosonic contributions $h_{i}\cdot\partial x(z_{i})$ can
contract either a plane wave factor leading to $h_{i}\cdot k_{j}\,
\langle\partial x(z_{i}) x(z_{k})\rangle= h_{i}\cdot
k_{j}\partial_{z_{i}}\ln\chi(z_{ij})$ (with $i\neq j$) or
contract a left moving $h_{j}\cdot\partial x(z_{j})$ (with $i\neq
j$) leading to $h_{i}\cdot h_{j}\partial^2_{z_{i}}\ln\chi(z_{ij})$
or a right moving $\bar h_{j}\cdot\bar\partial x(z_{j})$ (with
$i\neq j$) leading to $h_{i}\cdot \bar
h_{j}\partial_{z_{i}}\bar\partial_{z_{j}}\ln\chi(z_{ij})$.
The
bosonic propagator is given by
\eqn\newv{\eqalign{ \ln\chi(z) &={\pi\tau_2
\nu^2\over 2}- {1\over4} \ln\left|\sin(\pi z)\over \pi
\right|^2 - \sum_{m\geq1} \left({q^m\over 1-q^m} {\sin^2(m\pi
z)\over m}+c.c.\right) }}
where $q= \exp(2i\pi\tau)$.

The contractions between the fermions in the vertex operators
 are given by  $\langle \psi^\mu(z)\psi^\nu(0)\rangle_\alpha =\ap\,
\eta^{\mu\,\nu}\, S_\alpha(z)$
\eqn\ePsiPsi{
S_\alpha(z)= {\theta_{\alpha}(z|\tau)\over \theta_{\alpha}(0|\tau)}\, {\theta_{1}'(0|
\tau)\over \theta_{1}(z|\tau)}
}
for the even spin structures $\alpha=2,3,4$ and
\eqn\eSodd{
S_{1}(z|\tau)={ \theta'_{1}(z|\tau)\over \theta_{1}(z|\tau)}
}
for the odd spin structure.
Performing such contractions in eq.~\eAn\ for $T_{2n}\cdot F^n$ one gets
\eqn\eTn{\eqalign{
T_{2n}\cdot F^n&= \sum_{\sigma \in S_{n}\atop\sigma=(c_{1})\cdots (c_{k})}
{\rm tr}(F^{i_{c_{1}(1)}}\cdots F^{i_{c_{1}(l_{1})}})\cdots
{\rm tr}(F^{i_{c_{k}(n-l_{k}+1)}}\cdots F^{i_{c_{k}(n)}})\cr
&\hskip55pt\times G(z_{\sigma(1)}-z_{\sigma(2)},\cdots,z_{\sigma(n-1)}-z_{\sigma(n)})
}}
Hence $T_{2n}\cdot F^n$ is expressed as the sum of products of traces over the  decomposition of the
permutations
$\sigma$ of the $n$ indices over a product of cycles $c_{k}$ of length $l_{k}$.
Because ${\rm tr}(F)=0$ no cycle of length 1 can occur in the decomposition.
The function $G$ is expressed in terms of the fermionic propagators as
\eqn\eRId{
G(x_{1},\dots,x_{n}|\tau)= \sum_{\alpha=2,3,4} \, (-1)^{\alpha-1}\,
{\theta^4_{\alpha}(0|\tau)\over \eta^{12}(\tau)}\, \prod_{j=1}^{n} \, S_{\alpha}
(x_{j})
}
where $x_{1}+\cdots+x_{n}=0$.

The Jacobi identity insures that $G(x_{1},\cdots,x_{n})=0$ for $n\leq3$
and $G(x_{1},\cdots,x_{4})=1$.
In  the four-point amplitudes the only cycle decompositions of $\sigma\in S_{4}$
that
contribute are $\sigma=(1234)$ and $\sigma=(12)(34)$ and their cyclic
permutations,
giving rise to the famous $t_{8}F^4$ tensor.
Using an extension of the Fay trisequent formula one can explicitly evaluate
to all
orders the sum over the spin structure \refs{\BohrVanhove}.
The result for $n=5$  is given by
\eqn\eRfive{
G(x_{1},\dots,x_{5})= \sum_{j=1}^5 S_{1}(x_{j})
}
with $x_{1}+\cdots +x_{5}=0$.

\smallskip
The odd spin structure begins to contribute from $n\geq5$ points onwards
and takes the form
\eqn\eOneoL{\eqalign{
A^{odd}_{n}&={\kappa^2_{(10)}\over{ \alpha'}^{{D\over2}-n}}\,\int_{\cal F}{
d^2\tau\over \tau_{2}}
\int d^{10}\psi_{0}
d^{10}\bar\psi_{0}\, \tau_{2}^{n-5}\, \Gamma_{(10-D,10-D)}\times\cr
&
\prod_{i=1}^{n-1}\int {d^2{\bf z}_{i}\over \tau_{2}}\,
e^{\psi_{0}^\mu  (i k_{i}^\mu \theta_{i}+h_i^\mu)+c.c.} \,{\cal A}^{odd}_{n}
\,\prod_{1\leq i<j\leq n} |\chi(z_i-z_j)|^{-2\alpha' k_{i}\cdot k_{j}}
}}
Here $d^2{\bf z}=d^2z\, d\theta\, d\bar\theta$ is the measure of
integration over the positions of the insertion points of the vertex
operators in the $N=1$ world-sheet formalism. The amplitude receives
an odd spin structure contribution from $n\geq5$~\refs{\PeetersUB}
in ten dimensions.
For a toroidal compactification the number of fermionic zero modes will
not depend on the dimension because $\cN=8$ supersymmetries are preserved.
The integration over the fermionic zero modes is carried out using the rule
\eqn\ePsiint{
\int d^{10}\psi_{0}\, \psi_{0}^{m_{1}}\cdots \psi_{0}^{m_{10}}= \ap^5\, 10!\,
\epsilon_{10}^{m_{1}\cdots m_{10}}
}
For a compactification of the loop amplitude on a torus of dimension
$10-D$, we will have $D$ zero modes from the space-time part and
$10-D$ zero modes from the internal fermions along the torus
directions. For the case of amplitudes with only graviton
vertex operators it is not possible to saturate the fermionic zero
modes from the internal directions and the amplitude vanishes in $D<10$.

\subsec{The field theory limit}

In this paper we are interested in the low-energy limit $\ap\to0$ of
the $n$-graviton type~II string amplitude in $4\leq D\leq10$ dimensions. The
limit is achieved as in \refs{\GreenSW} and leads to the $\cN=8$
supergravity field theory amplitude evaluated in the dimensional
regularisation scheme.

Compactified on a $10-D$ dimensional square torus of typical size $R$ the string
amplitude described in the previous section takes the form
\eqn\eAgO{
A^{genus-1}={\kappa^2_{(10)}\over\ap^{{D\over2}-n}}\, \int_{\cal F} {d^2\tau\over \tau_2} \,\tau_2^{n-5}\,
F_n(\tau) \, \Gamma_{(10-D,10-D)}
}
where $F_n(\tau)$ is the integrand of the $n$-graviton amplitude
given in eq.~\eOneeL. Because we are interested in the
supergravity limit all the winding modes and Kaluza-Klein states will
be decoupled by taking the scaling limit $R\to0$ and $\ap/R\to 0$
\refs{\GreenYU,\GreenSW} (at one-loop the limit is not affected by
the issue raised in \refs{\GreenZZB}) with the result \eqn\eGG{
\lim_{\ap\to0\atop R\to0,\ap/R\to0}\Gamma_{(10-D,10-D)}\to
R^{5-{D\over2}}\, \tau_2^{5-{D\over2}} }

In the limit  $\ap\to0$ one has to take the string
proper time $\tau_{2}\to\infty$ so that $t=\alpha' \tau_{2}$
and the positions\foot{The differences between
the $\nu_{i}$ give the Feynman parameters of the field theory loop
amplitude once an ordering of the external leg has been chosen.}
of the vertex operators
$z_{i}=z^{(1)}_i+i\tau_2\nu_i$ with $\nu_i\in[0,1]$
stays finite.
As well there are some contributions from
colliding  several (two or more) vertex
operators, leading to the reducible contributions represented in~\figred.

\smallskip

Because of the vanishing of the one-, two-, and three-point amplitudes in
type~II superstring and $\cN=8$ supergravities, reducible contributions
can only appear from
$n\geq5$ graviton amplitudes and are constructed from boxes and higher
point amplitudes.

\smallskip

In the scaling limit $\ap\to0$, one has to take $\tau_2\to\infty$ and
$t=\ap\,\tau_2$ finite.  The bosonic and fermionic propagators have the
following limiting expressions
\eqn\eSinf{
S_1(z)\to G_F(\nu)\equiv \pi\, \sg(\nu) }
and
\eqn\ePinf{\eqalign{
\partial_{z}\ln \chi(z)&\to\dot G_B(\nu)\equiv {\pi\over2}\,\big( 2\nu -
{\rm sign}(\nu)\big)\cr
\partial^2_{z}\ln\chi(z)&\to\ddot G_B(\nu)\equiv -{\ap\pi\over4 \,t}\cr
\partial_{z}\bar\partial_{\bar z}\ln\chi(z)&\to\ddot G_B(\nu)\equiv -
{\ap\pi\over4 \,t}
}}
Because of the zero mode contributions from the coordinates $x^\mu(z)$
the second derivative of the bosonic propagator contribute to an
inverse power of the proper time. This
leads to a shift in the dimension of the
resulting field theory loop amplitude.

\smallskip

In this limit the fermionic contractions $T_{2n}\cdot F^n$ in
eq.~\eTn\ leads to
\eqn\eTninf{\eqalign{
 t_{2n}\cdot F^n&\equiv\lim_{\ap\to0} T_{2n}\cdot F^n \cr
&=\sum_{\sigma\in S_n}\, {\rm tr}(F^{i_{c_{1}(1)}}\cdots F^{i_{c_{1}(l_{1})}})
\cdots
{\rm tr}(F^{i_{c_{k}(n-l_{k}+1)}}\cdots F^{i_{c_{k}(n)}})\cr
&\times G^\infty(z_{\sigma(1)}-z_{\sigma(2)},\cdots,z_{\sigma(n-1)}-z_{\sigma(n)})
}}
where $G^\infty(x_1,\cdots,x_n)=0$ for $n\leq3$, $G^\infty(x_1,\cdots,x_4)=1$ and
\eqn\eGFfive{
G^\infty(x_1,\cdots,x_5)=\sum_{i=1}^5 \, \sg(x_i)
}

\smallskip

In the field theory limit the factor from the contractions between
the plane waves approximates to

\eqn\eChi{
\left\langle\prod_{1\leq i<j\leq n} e^{ik_i\cdot x(z_i)}\right\rangle=\prod_{1
\leq i<j\leq n} \, \chi(z_i-z_j)^{-\alpha' k_{i}\cdot k_{j}}
\to \prod_{1\leq i<j\leq n} \,\exp\left(-\pi\,t\,Q_{n}\right)
}
with $Q_n$ defined in eq.~\eQn.
Depending on the number of first and second derivatives of the bosonic
propagators and the number of fermionic propagators, one gets
that the field theory one-loop integrals are given by
\eqn\eAgOO{\eqalign{
M_n &= \int_0^1 {dt\over t} \, t^{m+n-{D\over2}}\, \prod_{i=1}^{n-1}
\int_0^1 d\nu_i\, \nu_1\cdots \nu_k\, e^{-\pi \, t \,Q_n}\cr
&= \pi^{{D\over2}-m-n}\, \Gamma(m+n-{D\over2})\,  \prod_{i=1}^{n-1}
\int_0^1 d\nu_i\, \nu_1\cdots \nu_k\, Q_n^{{D\over2}-m-n}
}}
This is the expression for the $n$-point integrals $I_n^{(D+2m)}[\nu_1
\cdots \nu_n]$ summed over all orderings of the external legs.

\listrefs \bye